\documentclass[12pt,showpacs,aps,preprintnumbers,prb,amsmath,amssymb]{revtex4}
\usepackage{graphicx}
\usepackage{epsfig,color}
\def\ThPtGe{ThPt$_4$Ge$_{12}$}
\begin{document}
%%%%%%%%%%%%%%%
%\draft

\title{Possible anisotropic superconducting pairing in cubic ThPt$_4$Ge$_{12}$}

\author{V. H. Tran $^1$, A. D. Hillier $^2$, D. T. Adroja $^2$, D. Kaczorowski $^1$}
\address{$^1$Institute of Low Temperature and Structure Research, Polish Academy of Sciences, P. O. Box 1410, 50-950 Wroc\l aw, Poland\\
$^2$ISIS Facility, Rutherford Appleton Laboratory, Chilton, Oxfordshire OX11 0QX, United Kingdom\\
}

\begin{abstract}
Transverse-field muon-spin-rotation measurements have been carried out for polycrystalline \ThPtGe. The magnetic penetration depth $\lambda$ and the superconducting coherence length $\xi$ in the vortex state of this compound were found to be 112(5) nm and 32(2) nm, respectively. We have estimated the effective mass of the quasiparticles $m^* \approx 4.6\times m_e$, the superfluid carrier density $n_s \approx 1.02 \times 10^{28}$ carriers/m$^{3}$, and the zero-temperature superconducting gap 0.67 meV, corresponding to the ratio: 2$\Delta(0)/k_BT_c$ = 3.76.   We found markable difference between the temperature dependence of the vortex state muon-spin relaxation rate ZFC- and FC-$\sigma_s(T)$ below the irreversibility temperature $T_{ir} \sim $ 2.5 K. Linear field dependence of $\lambda(H)$, small ratio $T_{ir}/T_c$, and power law behavior of the temperature dependencies of  $\lambda(T)$ seem to be consistent with anisotropic superconducting pairing in the compound studied. The analysis of correlation between the superconducting transition temperature and the effective Fermi temperature within the Uemura classification scheme reveals that \ThPtGe~belongs to the same class of exotic superconductors.
\end{abstract}
\pacs{74.70.Dd; 74.25.Op;74.20.Rp;76.75.+i}
\maketitle

\newpage

\section{Introduction}
\par Recently, two research groups have independently reported discovery of a novel class of ternary compounds with the filled skutterudite structure MPt$_4$Ge$_{12}$, where M=Sr, Ba, La, Ce, Pr, Nd, and Eu.\cite{Bauer07,Gumeniuk} The compounds with Sr and Ba have been found to be superconducting below $T_c$ = 5.10 and 5.35 K,\cite{Bauer07} respectively, whereas those with La and Pr have been found with $T_c$ = 8.3 and 7.9 K,\cite{Gumeniuk} respectively. We also reported on the formation and the superconducting properties of another Ge-based skutterudite \ThPtGe~with $T_c$ = 4.62 K, which is the actinoid bearing representative of the MPt$_4$Ge$_{12}$ series.\cite{KT08} The compound can be characterized by the Ginzburg-Landau coherence length of $\xi \sim$ 35 nm, the penetration depth of $\lambda \sim$ 150 nm, electronic mean free path $\l_e \sim$ 435 nm, and the upper critical field $\mu_0 H_{c2}$ = 0.29 T. A large ratio $\l_e/\xi$ signifies \ThPtGe~to be in the clean limit. Subsequently, the superconductivity in this material has been confirmed by Bauer \emph{et al}.\cite{Bauer08} Thus, like the compounds with M = Sr, Ba, La, and Pr, the \ThPtGe~compound exhibits type-II superconductivity at low temperatures. Owing to the fact that \ThPtGe~and other MPt$_4$Ge$_{12}$ exhibit similar values of the critical temperatures, which range from $\sim$ 5 to $\sim $ 8 K, one expects a minor role of the filler M element, but a dominating contribution to the superconducting properties from interactions within the Pt-Ge framework. In natural consequence, it is believed that superconductivity in these skutterudites is most likely mediated by phonons. However, some experimental data on \ThPtGe,\cite{KT08} indicated possibility of an unconventional character of the electron pairing, as in the superconducting state both the temperature dependence of the electronic specific heat $C_{es}$ and the magnetic field variation of the Sommerfeld coefficient exhibit some behavior characteristic for the point-like node superconductivity.  For \ThPtGe, $^{195}Pt-$NMR experiments and electronic band structure calculations,\cite{Tran09b} have suggested that the physical properties of this compound may be anisotropic. Furthermore, the Hall effect and thermoelectrical power data have revealed the existence of both electrons and holes,\cite{Tran09a} manifesting different behavior from those of conventional BCS ordinary superconductors, which usually exhibit one-band conduction. The unclear mechanism of electron pairing in \ThPtGe~shows that further experimental and theoretical studies of this compound are needed.
\par In this paper, we report a study of the vortex state in \ThPtGe~by means of transverse-field muon-spin rotation (TF-$\mu$SR) technique, which is very sensitive measurement of the microscopic field distribution inside type-II superconductors.\cite{Amato,Sonier1,Kadono,Sonier2} From the measurements, we have estimated the magnetic penetration depth $\lambda$ and the coherence length $\xi$, and then we have derived an effective mass of the quasiparticles $m^*$ and superconducting carrier density $n_s$. Applying zero-field cooling and field-cooling sample modes we have investigated the flux pining phenomenon in \ThPtGe, and have found the irreversibility temperature $T_{ir}$. Interestingly, the temperature dependencies of the magnetic penetration depth $\lambda(T)$ and field dependence of $\lambda(H)$ exhibit deviations from those expected for the isotropic s-wave superconductors. A small ratio $T_{ir}/T_c$ and close relationship of \ThPtGe~to unconventional superconductors within the Uemura classification scheme,\cite{Uemura} are further evidences of unusual behavior of the investigated superconductor. Very recently the transverse-field (TF)-muon spin rotation experiments for PrPt$_4$Ge$_{12}$ have established the presence of point-like nodes in the superconducting energy gap.\cite{Maisuradze}

\section{Experimental details}
\par Polycrystalline sample of about 5 g used in this study was prepared and characterized as reported previously.\cite{KT08} $\mu SR$ measurements were carried out using MuSR spectrometer installed at the ISIS Facility of the Rutherford Appleton Laboratory, Chilton, UK, where pulses of muons are produced every 20 ms with FWHMs of 70 ns. For the measurements, pulverized sample was mixed with GE varnish and glued onto a high purity silver ($>$4N) holder of 30 mm diameter and 1 mm thick. We have carried out the measurements at temperatures ranging from 0.3 to 5 K in magnetic field up to 0.06 T, applying two different modes of cooling the sample: in zero magnetic field (ZFC) and applied magnetic field (FC).

\section{Results and discussion}
\par
\begin{figure}[h]
\includegraphics[scale=1]{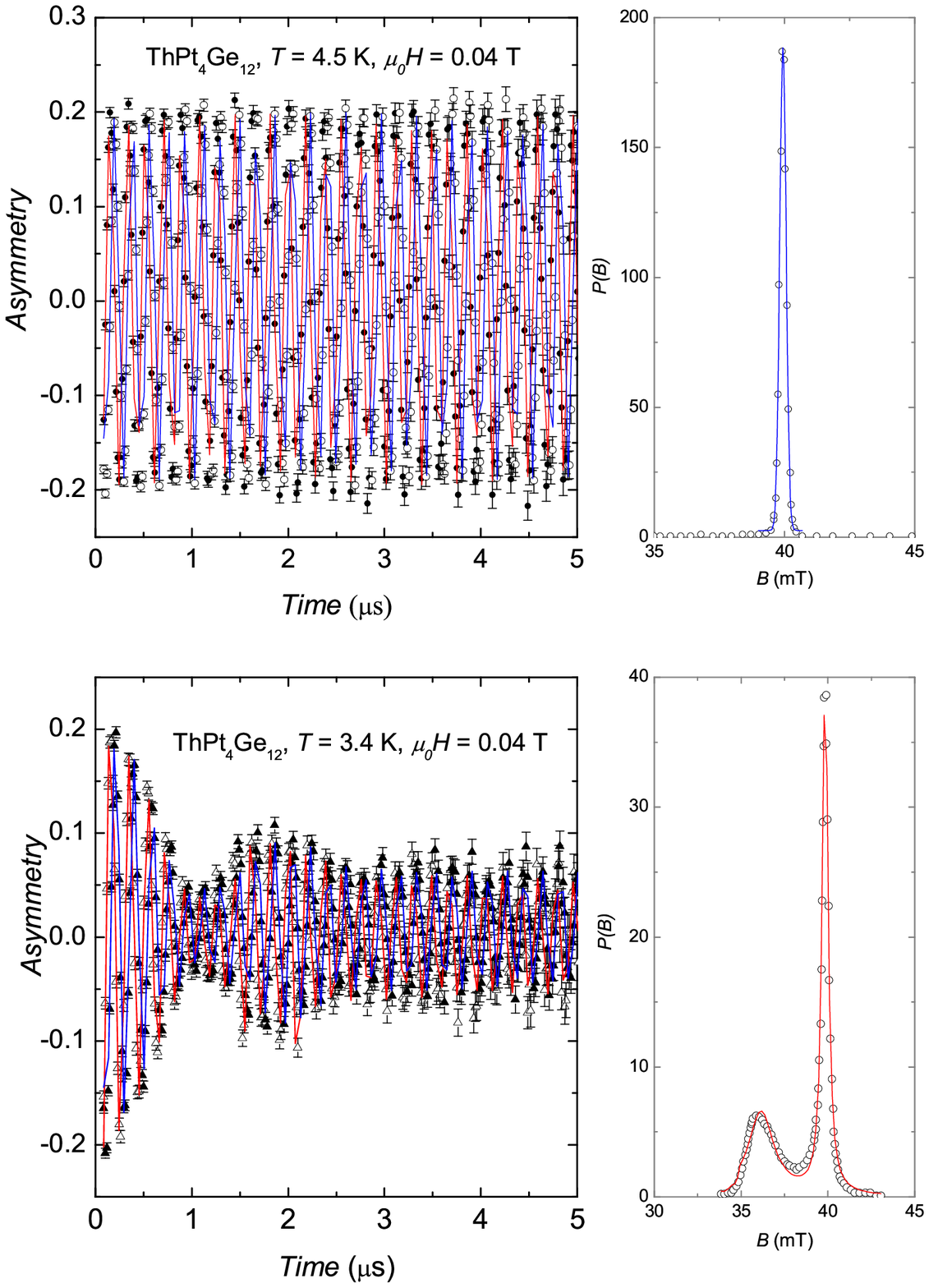}
\caption{\label{fig:Fig_1} (color online) Transverse-field $\mu$SR time spectra and internal magnetic field distributions of \ThPtGe~taken above $T_c$ (upper panel) and below $T_c$ (lower panel) in applied magnetic of 0.04 T. The solid lines represent the best fit with eq. \ref{eq:G1}. The dashed lines are the fits to Gaussian functions. Open and close symbols in TF-$\mu$SR time spectra denote real and and imaginary components, respectively.}
\end{figure}

 Typical TF-$\mu SR$ time spectra for \ThPtGe, taken above $T_c$ and below $T_c$, are shown in Fig. \ref{fig:Fig_1}. Comparison of the spectra indicates clear difference in the relaxation rate $\sigma$ above and below $T_c$. This difference is related to the change in magnetic field distribution inside the sample, which can be obtained using the algorithm of fast Fourier transformation, and it is displayed on the right- hand side of the spectra. In the normal state, the field distribution corresponds to the applied external magnetic field of $B_0$ = 40 mT and the Gaussian-like peak with some broadening is attributed to the nuclear magnetic moments. Below $T_c$,  due to the formation of the flux line lattice state associated with the superconductivity, the magnetic field  distribution becomes inhomogeneuos and two Gaussian components are observed. The larger component at  $B_0$ ascribes the field distribution of nonsuperconducting signal, whereas the smaller peak corresponds to the second moment of the field distribution $<\Delta B^2> = <[B(r)-B_0]^2> $, arising due to the formation of the flux line lattice.
To elucidate details of the vortex state we have analyzed the data using the expression:
\begin{equation}
A_{TF}(t) = A_0exp(-\frac{1}{2}(\sigma_s t)^2)exp(-\frac{1}{2}(\sigma_n t)^2)cos(\omega t+\Phi_1)+A_{bg}cos(\omega_{bg} t+\Phi_2),
\label{eq:G1}
\end{equation}
where the first term is the contribution from the sample and the second term from the background. $A_0$ and $A_{bg}$, $\omega = 2\pi \gamma_\mu B_r$ and $\omega_{bg}= 2\pi \gamma_\mu B_0$, $\Phi$ and $\Phi_{bg}$, are the initial asymmetries, muon precession frequencies, and phase angles for the sample and background, respectively. $\sigma_s$ and $\sigma_n$ are the muon-spin relaxation rates in the superconducting and normal state, respectively. $\gamma_\mu/(2\pi)$ = 135.53 MHz/T is the gyromagnetic ratio of the muon.

\begin{figure}[h]
\includegraphics[width=1\textwidth]{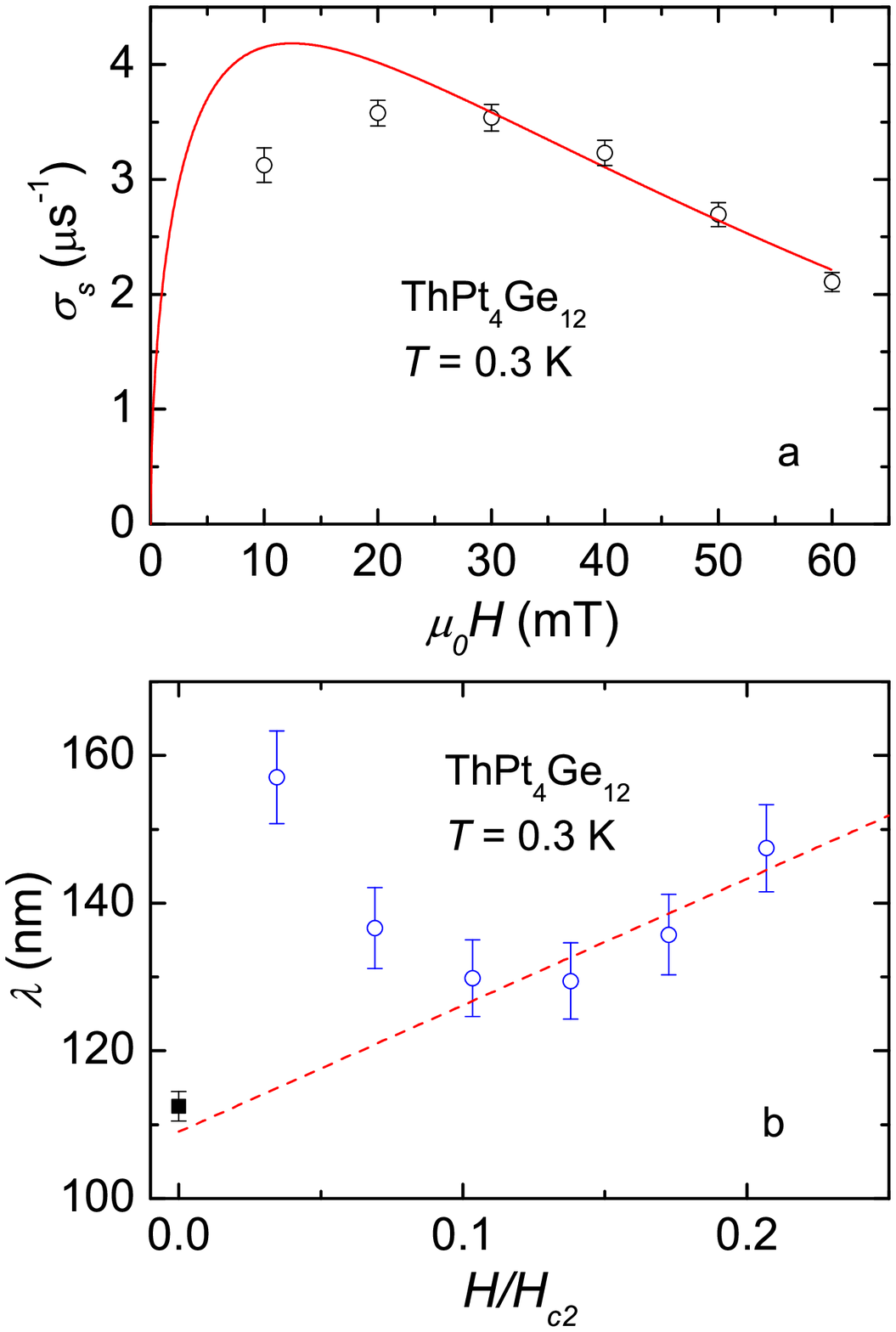}
\caption{\label{fig:Fig_2} (color online)  a) Field dependence of the TF-muon spin-relaxation rate $\sigma_s$ of \ThPtGe~at 0.3 K. The solid line is the fit of eq. \ref{eq:S1} to the experimental data. b) The magnetic penetration depth $\lambda$ vs. reduced field $H/H_{c2}$. The closed square is the fitted value with  eq. \ref{eq:S1}. The dashed line indicates linear dependence of  $\lambda(H/H_{c2})$ for 0.1 $< H/H_{c2} < $0.2 with the slope of 1.6. }
\end{figure}

\begin{figure}[h]
\includegraphics[width=1\textwidth]{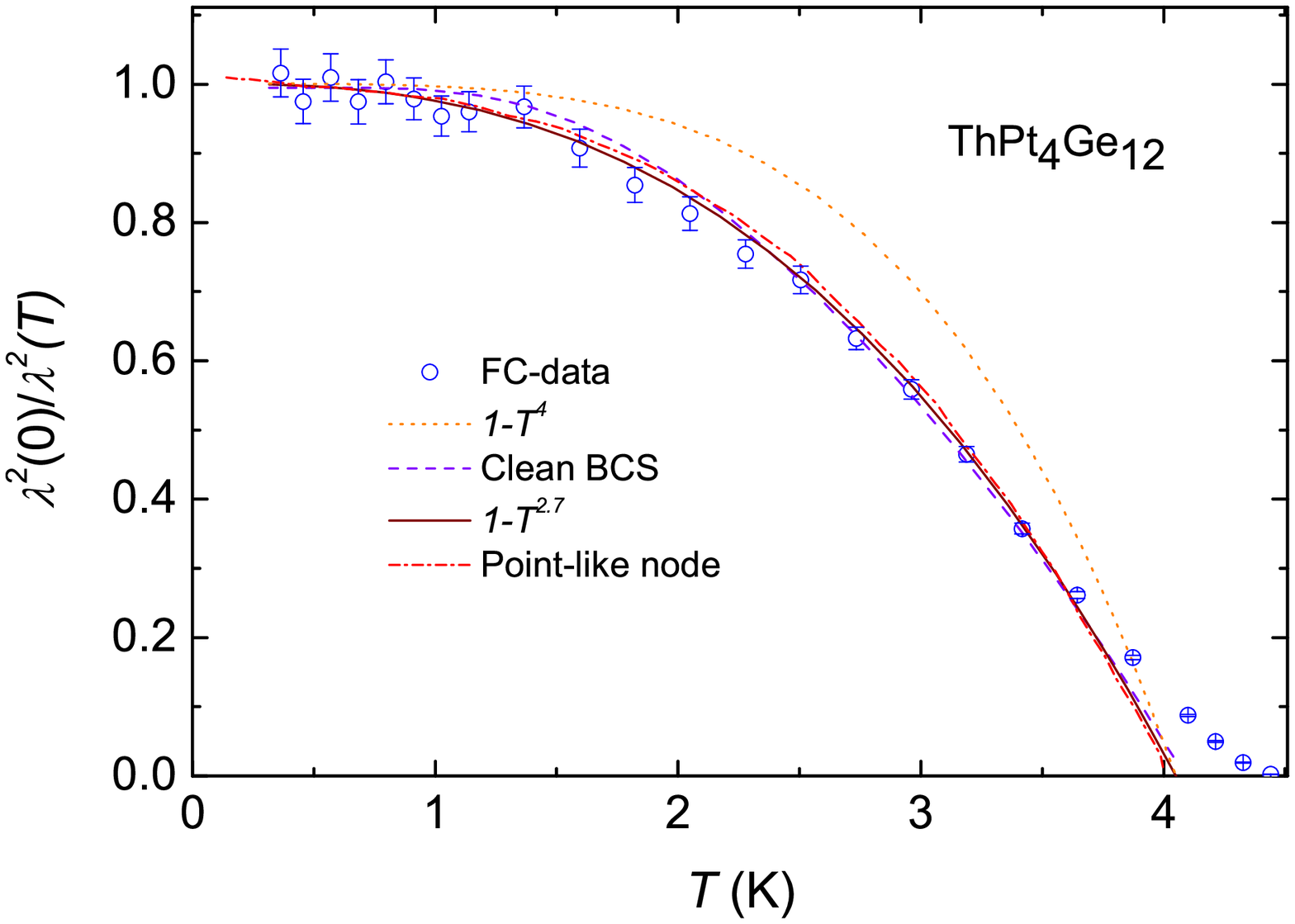}
\caption{\label{fig:Fig_3} (color online) Temperature dependence of the normalized superfluid density of \ThPtGe~in an applied magnetic field of 0.04 T. The lines are the fits discussed in the text. }
\end{figure}

\begin{figure}[h]
\includegraphics[width=1\textwidth]{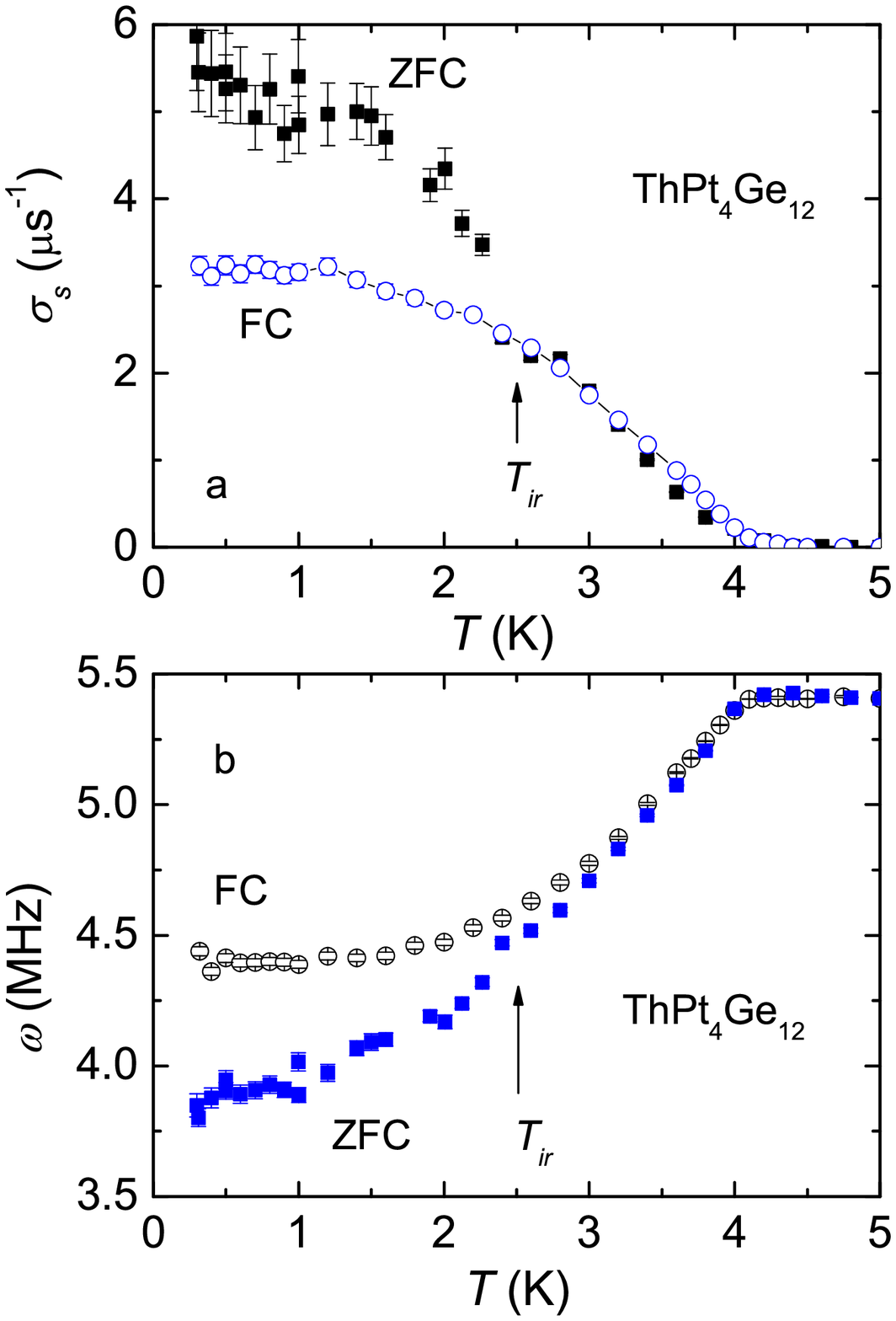}
\caption{\label{fig:Fig_4} (color online)
Temperature dependence of the TF-muon a) spin-relaxation rate $\sigma_s$ and b) muon precession frequency $\omega$ from the ZFC and FC measurements in a field of 0.04 T. }
\end{figure}

Fits to the TF-$\mu SR$ time spectra, shown in Fig. \ref{fig:Fig_1} as the solid lines, have been obtained as follows. First we have fitted the normal state data, both the real and imaginary components, keeping $\sigma_s$ = 0. In this way the values of $A_0$, $\sigma_n$, $\Phi_1$, $\Phi_2$, $A_{bg}$, $\omega$ and $\omega_{bg}$ have been determined. For the data below $T_c$, we have kept $A_0$, $A_{bg}$, $\sigma_n$ as the same values as in the normal state, and from the fits we have determined the parameters $\sigma_s$, $\omega$, $\omega_{bg}$, $\Phi_1$ and $\Phi_2$. The obtained values of the parameters $\sigma_s$ and $\omega$ as functions of the magnetic field strength and the temperature are shown in Figs. \ref{fig:Fig_2} - \ref{fig:Fig_4} and will be discussed below.
\par Fig. \ref{fig:Fig_2} a shows the field dependence of the muon-spin relaxation rate $\sigma_s$ in \ThPtGe~obtained at 0.3 K. We see that $\sigma_s (H)$ depends strongly on applied fields. A similar behavior has been observed in PrOs$_4$Sb$_{12}$,\cite{MacLaughlin} KOs$_2$O$_6$,\cite{Koda} and PrRu$_4$Sb$_{12}$.\cite{Androja05} From the field dependence of the muon relaxation rate, we have evaluated the magnetic penetration depth $\lambda$ and the coherence length $\xi$ using the modified London equation:\cite{Brandt}
\begin{equation}
 \sigma_s (H) =(\frac{\gamma_\mu}{ \sqrt{2}})H \sum_{h,k} [\frac{exp(-\xi^2 q_{h,k}^2)}{[1+q_{h,k}^2 \lambda^2/(1-H/H_{c2})^2]^2}]^{1/2},
 \label{eq:S1}
 \end{equation}
where $q_{h,k} \neq 0$ is the lattice sum over the hexagonal flux line lattice. The best fit (see the solid line) gave $\lambda$ = 112(5) nm and $\xi$ = 32(2) nm, which corroborate the values derived from the specific heat measurements with $\lambda \approx$  150 nm and  $\xi \approx$ 35 nm, respectively.\cite{KT08}
\par
Generally, the magnetic penetration depth is related to the superconducting carrier density $n_s$, effective mass $m^*$, coherence length $\xi$  and the mean free path  $l_e$. In clean limit, $\xi << l_e$, $\lambda(0)$ is given as:
\begin{equation}
\lambda (0) = (\varepsilon_0m^*c^2/n_sq^2)^{1/2},
\label{eq:lambda}
\end{equation}
where \emph{c} and \emph{q} have the usual meaning. The estimation of $n_s$ and $m^*$ is impossible
using the $\mu$SR data alone, and hence it is helpful assuming that all the normal state carriers contribute to the superconductivity, i.e.,  the coefficient of the electronic specific heat is related to $m^*$ and $n_s$ via the relation:
\begin{equation}
\gamma = \frac{m^*(3\pi^2n_s)^{1/3}k_B^2}{3\hbar^2}
\label{eg:Specific}
\end{equation}
Taking $\lambda$ = 112 nm and $\gamma(0.4 {\rm K}, 0.5 {\rm T})$ = 42 mJ mol$^{-1}$ K$^{-2}$ from Ref. \onlinecite{KT08}, and combining Eqs. \ref{eq:lambda} and \ref{eg:Specific}, one evaluates $m^*$ = 4.6$m_e$ and $n_s$ = 1.02$\times$10$^{28}$ carriers/m$^{3}$. The obtained values substantiate that \ThPtGe~is a good metal with enhanced mass carriers.

\par It is known that the field dependence of the magnetic penetration depth can provide information on the degree of anisotropy of the superconducting order parameter.\cite{Kadono} In fact, the $\lambda(H)$-dependence is entirely different for conventional BCS-type superconductors and unconventional superconductors with anisotropic energy gaps. Therefore, it is tempting to consider the $\lambda(H)$ dependence for \ThPtGe. It can be derived from the relation:\cite{Brandt2}
\begin{equation}
\sigma_s[\mu s^{-1}] = 4.83\times10^4(1-H/H_{c2})[1+1.21(1-\sqrt{H/H_{c2}})^3]\lambda^{-2}
\label{Eq_S2}
\end{equation}
Note that this equation holds for $\kappa \geq$ 5 and 0.25/$\kappa^{1.3} \leq H/H_{c2} \leq 1 $. Because ThPt$_4$Ge$_{12}$ has $\kappa$ = 3.5 then the description  $\lambda(H)$ with Eq. \ref{Eq_S2} may have an error larger than 5 \%.
 Taking temperature dependence of $H_{c2}$,\cite{KT08} and using Eq. \ref{Eq_S2} we evaluated $\lambda$ and plot the $\lambda(H/H_{c2})$-dependence in Fig. \ref{fig:Fig_2} b. Apparently, in a similar manner as found for various unconventional superconducting materials, $\lambda$ of \ThPtGe~is increased almost linearly with the field in a certain range of the strength of applied field. A linear fit of the $\lambda(H/H_{c2})$ data in the field interval 0.1 $< H/H_{c2} < $0.2 to the function $\lambda = \lambda(0)(1+\eta H/H_{c2})$ yields $\eta$ = 1.6(1). Here, one should recall that unconventional superconductors $\eta$ takes values between 1 and 6, in contrast to the case of isotropic gap BCS-type superconductors for which it is equal to zero. The $\eta$ value determined for \ThPtGe~is  close to that found for isotropic superconductors KOs$_2$O$_6$ (n = 2.58),\cite{Koda}  and  NbSe$_2$ ($\eta$ = 1.61),\cite{Sonier97} but smaller than that derived for the high-$T_c$ superconductor with d-wave pairing YBa$_2$Cu$_3$O$_{6.6}$, ($\eta \sim$ 6).\cite{Sonier97b} This comparison suggests possible anisotropic pairing in \ThPtGe.
\par Fig. \ref{fig:Fig_3} shows the normalized superfluid density $\lambda^{-2}(T)/\lambda^{-2}(0)$  in the FC regime for \ThPtGe~in an applied magnetic field of 0.04 T versus temperature $T$. The temperature dependence of the magnetic penetration depth $\lambda^{-2}(T)$ has been inferred data using Eq. (\ref{Eq_S2}) and taking the $H_{c2}(T)$,\cite{KT08} and $\sigma(T)$ data. According to the Gorter-Casimir two-fluid model for s-wave isotropic gap superconductors, $\lambda^{-2}(T)/\lambda^{-2}(0)$ is proportional to $(1-(T/T_c)^4)$. As shown in Fig. \ref{fig:Fig_3}, the TF-data of \ThPtGe~exhibit significant deviation from this relation (dotted line). This observation indicates that \ThPtGe~might be not an isotropic BCS superconductor. It turns out that the experimental data can be fitted to the relation:
\begin{equation}
\lambda^{-2}(T)/\lambda^{-2}(0) = 1-(T/T_c)^n
\end{equation}
with $\lambda(0)$ = 130(5) nm, \emph{n} = 2.7(1), $T_c$ = 4.1 K (note the solid line). We would mention that the observed \emph{n}-value is close to \emph{n} = 2, predicted for nodal superconductors.\cite{Annett} Similar reducing in \emph{n}-value has been observed in KOs$_2$O$_6$ (n = 2.39),\cite{Koda} and in PrRu$_4$Sb$_{12}$ (\emph{n} =1.44).\cite{Androja05}
\par Since \ThPtGe~is in the clean limit, the relation:
\begin{equation}
\lambda^{-2}(T)/\lambda^{-2}(0) = 1-2\int_{\Delta(T)}^\infty(-\frac{\partial f}{\partial \varepsilon})\frac{\varepsilon}{\sqrt{\varepsilon^2-\Delta(T)^2}}d\varepsilon,
\label{CL}
\end{equation}
should be tested for the experimental data. In the above equation, $f = [1+exp(\varepsilon/k_BT)]^{-1}$ is the Fermi function, $\Delta(0)$ is the zero-temperature value of the superconducting gap, and $\Delta(T) = \Delta(0){\rm tanh}(1.82[1.018(T_c/T-1)]^{0.51}$),\cite{carrington} represents  the temperature dependence of the gap. Fitting the experimental data with $\lambda(0)$ = 130 nm, taken equal to the extrapolated value at \emph{T} = 0 one obtains $T_c$ = 4.1 K and $\Delta(0)$ = 0.67 meV. These values lead to the ratio: 2$\Delta(0)/k_BT_c$ = 3.76, which is slightly larger than the weak-coupling BCS value of 3.52. Within the experimental error limit, the fit with eq. \ref{CL} is quite good for $T >$ 2.5 K but becomes little ambiguous for the temperature range 0.8 - 2.2 K (see the dashed line).
\par To check the possibility of unconventional superconducting order parameters in \ThPtGe, the magnetic penetration depth data are fitted to point-node model, because this model predicted the proportionality of the electronic specific heat $C_e \sim T^3$,\cite{Bash} as that we have previously found for \ThPtGe.  We used the following equations in evaluating the $\lambda(T)$ data:\cite{Prorozov}
\begin{equation}
\lambda^{-2}(T)_{aa}/\lambda^{-2}(0)_{aa}=1-\frac{3}{4T}\int_0^1(1-z^2)\int_0^{2\pi}(_{sin^2\varphi}^{cos^2\varphi})\int_0^\infty cosh^{-2}(\frac{\sqrt{\varepsilon^2+\Delta^2(z)}}{2T})d\varepsilon d\varphi dz
\end{equation}
\begin{equation}
\lambda^{-2}(T)_{cc}/\lambda^{-2}(0)_{cc}=1-\frac{3}{2T}\int_0^1z^2\int_0^{2\pi}{cos^2\varphi}\int_0^\infty cosh^{-2}(\frac{\sqrt{\varepsilon^2+\Delta^2(z)}}{2T})d\varepsilon d\varphi dz
\end{equation}
where $z = {\rm cos} \theta$, $\theta$ and $\varphi$ are the polar and the azimuthal coordinates in the k-space. In fitting we assumed $\Delta(z) =\Delta(T){\rm sin}\theta$ for the point node and the theoretical values are the average of the partial superfluid densities. The subscripts \emph{aa} and \emph{cc} indicate principal axis. The fit of the data to this point-node model, shown as the dashed line in Fig. \ref{fig:Fig_3}, is very good for $T <$ 1.5 K but little poor for the temperature range 1.5 - 3.0 K. We would like to note that $T_c$ deduced from the $\mu$SR measurements is fairly consistent with that extrapolated from $H_{c2}(T)$-dependence for a field of 0.04 T ($\sim$ 4.0(1)). \cite{KT08} Moreover, examining $\lambda(T)$  in Fig. \ref{fig:Fig_3}, we see that the $\lambda(T)$-curve has negative curvature around $T_c$. This behavior may indicate that the electronic properties in the vortex state of \ThPtGe~differ from those of isotropic BCS superconductors. Such a curvature of  $\lambda(T)$ was previously found for anisotropic superconductors YBa$_2$Cu$_3$O$_{7-\delta}$,\cite{Mao} and predicted theoretically for anisotropic p- or d-wave superconductors.\cite{PC}
\par In Fig. \ref{fig:Fig_4} a we show the temperature dependencies of the TF-muon spin-relaxation rate $\sigma_s$ taken in the ZFC and FC modes. These data point to the existence of distinct irreversibility with characteristic temperature  $T_{ir} \sim $ 2.5 K and below which $\sigma_{s, ZFC}$ being larger than $\sigma_{s, FC}$. The enhancement of  $\sigma_{s, ZFC}(T)$ at temperatures below $T_{ir}$ implies that there is an additional inhomogeneity of local-field distribution, which is related to flux pinning in the superconducting state. In other words,  when the flux is strongly pinned below $T_{ir}$, the flux vortex in the ZFC mode avoids the formation of the equillibrium flux vortex lattice. In contrast, at temperatures higher than $T_{ir}$, where the thermal fluctuation of the flux vortices overcomes the flux pinning to form the equillibrium flux lattice, the the field distribution in the ZFC regime is the same as that in the FC mode. This means that the temperature $T_{ir}$ is a result of the competition between the thermal fluctuation and the pinning of vortices. From the experimental values of the second moment of the field distribution, we have calculated the muon precession frequencies $\omega$ (Fig. \ref{fig:Fig_4} b). In a similar manner as in the case of the relaxation rate, the irreversible effect appears in the $\omega(T)$-curves below $T_{ir} \sim$ 2.5 K, namely $\omega$ measured in the FC mode is larger than that obtained in the ZFC mode. We may recall that $\mu$SR studies of flux pining phenomena for numbers of superconductors,\cite{Le,Wu} have shown that there is a relationship between the ratio $T_{ir}/T_c$ and the anisotropy of superconducting characteristics, namely the more anisotropic superconductor has a lower value of this ratio. Since the observed ratio of $T_{ir}/T_c \sim $ 0.54 is rather small, the superconducting gap in \ThPtGe~is likely anisotropic.

\begin{figure}[h]
\includegraphics[width=1\textwidth]{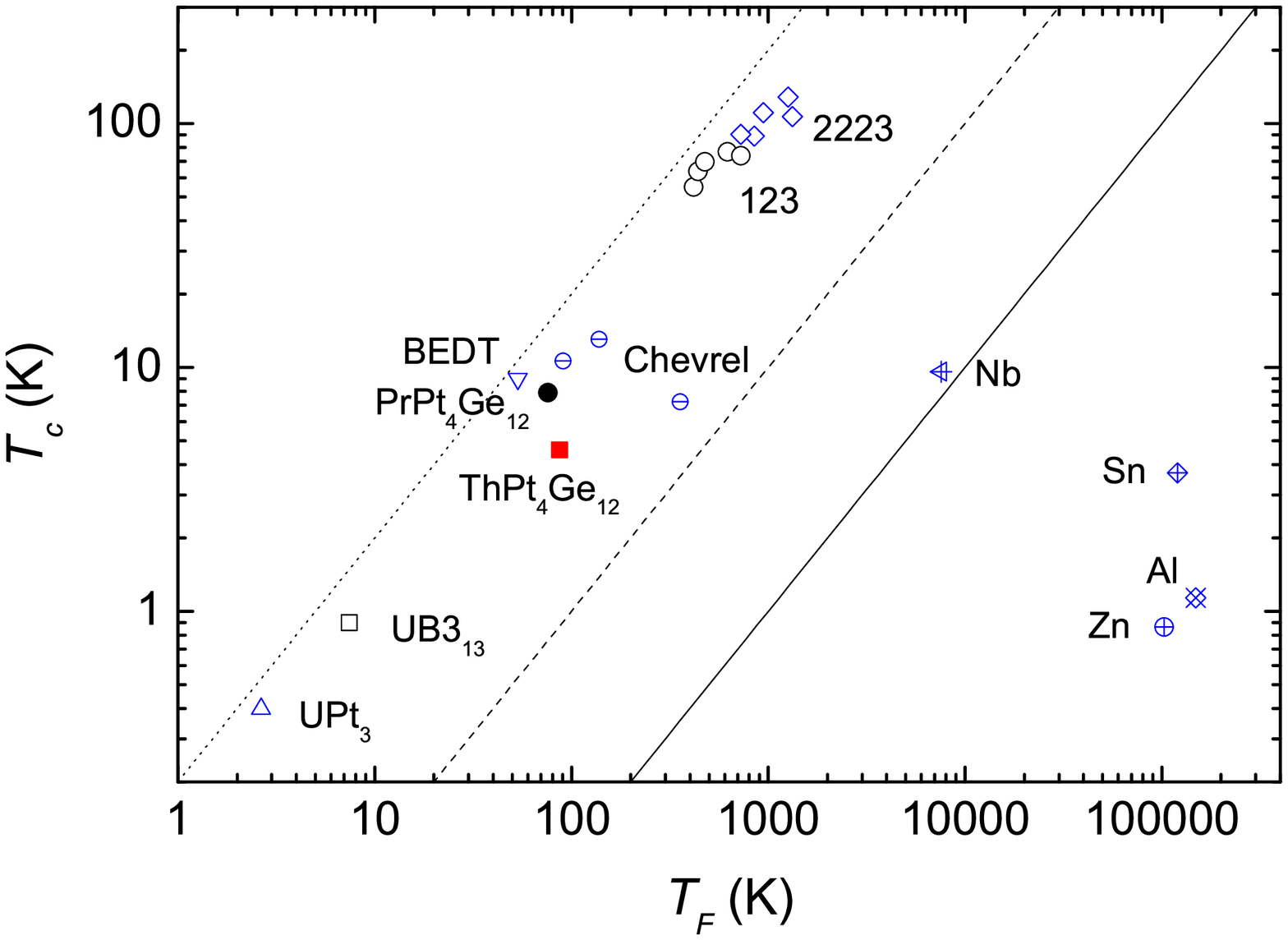}
\caption{\label{fig:Fig_5} (color online) Uemura plot, $T_c$ vs. $T_F$, for some selected superconductors including unconventional, isotropic s-wave,\cite{Uemura} and studied here \ThPtGe. The dotted and dashed lines 0.01 $ <T_c/T_F < $ 0.2 outline positions of unconventional superconductors from conventional BCS superconductors with solid line $T_c/T_F <$ 0.001.}
\end{figure}
\par Let us now discuss data of \ThPtGe~within the frameworks of the Uemura classification scheme.\cite{Uemura} According to the author, there exists a close relationship between the superconducting transition temperature $T_c$ and the effective Fermi temperature $T_F$, which can be determined from the $\mu$SR and heat capacity data using the formula $T_F = 730 \sigma^{3/4}\gamma^{-1/4}$, where $T_F$ in K, $\sigma$ in $\mu s^{-1}$, and $\gamma$ in mJ K$^{-2}$cm$^{-3}$. Interestingly, Uemura et al. was able to distinguish HTc, heavy-fermion and some exotic superconductors from the conventional BCS superconductors like Nb, Sn and Al. The first class of superconductors should fall in the range 0.01 $< T_c/T_F < $ 0.1, while the latter class of superconductors have $T_c/T_F < $ 0.001.\cite{Uemura} We re-plot in Fig. \ref{fig:Fig_5} the correlation between the superconducting temperature $T_c$ and the Fermi temperature $T_F$ for some selected superconductors, including the data for PrPt$_4$Ge$_{12}$ (calculated from data  of Ref. \onlinecite{Maisuradze}) and the data for \ThPtGe.  In Fig. \ref{fig:Fig_5} we have drawn two lines with the ratio $T_c/T_F$  of  0.2 and 0.01, intending to show the area in which unconventional superconductors are situated. For \ThPtGe~ the ratio $T/T_c$ = 0.053, much larger than 0.001, suggests that the pairing mechanism in this compound is different than that in isotropic s-wave superconductors but could be similar to that in exotic superconductors.

\section{Conclusions}
\par We have investigated the vortex state in ThPt$_4$Ge$_{12}$ by performing the transverse-field muon-spin-rotation in magnetic fields up to 0.06 T upon cooling the sample in zero and finite field (the ZFC and FC modes). The difference between the ZFC and FC data has been found to occur below the irreversibility temperature $T_{ir} \sim $ 2.5 K. From the field dependence of the vortex state muon-spin relaxation rate $\sigma_s(H)$, we have estimated the magnetic penetration depth $\lambda \sim$  112(5) nm, the coherence length $\xi \sim$ 32(2) nm. Combining the $\mu$ and specific heat data we have evaluated the effective mass of the quasiparticles $m^* \approx 4.6\times m_e$ and the superfluid carrier density $n_s \approx 1.02 \times 10^{28}$ carriers/m$^{3}$, confirming that \ThPtGe~is good metal with enhanced mass carriers. We have analyzed the temperature dependence of the superfluid density using various superconducting symmetry order parameters. Within the BCS clean limit approach, the fit of the experimental data revealed zero-temperature superconducting gap $\Delta(0)$ = 0.67 meV corresponding to the ratio: 2$\Delta(0)/k_BT_c$ = 3.76.  The phenomenological two-fluid model with the exponent of 2.7 and as well as the point-node model describe reasonably the $\lambda^{-2}(T)/\lambda^{-2}(0)$ dependence. Furthermore, we have observed the linear field dependence of $\lambda$ and the small ratio $T_{ir}/T_c$, which together with the $\lambda(T)$ behavior around $T_c$ might be taken as an evidence for an anisotropic order parameter. Comparison of the data with other superconductors in terms of the Uemura classification scheme suggests that \ThPtGe~and PrPt$_4$Ge$_{12}$ may share the pairing mechanism similar to that of exotic superconductors.
\par
%\section{Acknowledgments}
\par The authors are grateful to J. Sonier and A. Karen for an interesting discussion. The work was supported by the Polish Ministry of Science and Higher Education through the research grants N202 082/0449 and N202 116 32/3270. The measurement of $\mu$SR at ISIS is made possible
due to support from EU within NMI3 programme.

\end{document}